\documentclass[12pt]{article}
\usepackage{epsfig,amssymb,latexsym,amsmath,amsthm,fancyhdr}
\usepackage{graphics,psfrag,longtable}
\newcommand{\vp}{\mathbf{p}}
\newcommand{\bl}{\begin{aligned}}
\newcommand{\el}{\end{aligned}}

\newcommand{\la}{\langle}
\newcommand{\ra}{\rangle}

\newcommand{\vq}{{\mathbf{q}}}

\newcommand{\vk}{{\mathbf{k}}}

\newcommand{\be}{\begin{equation}}
\newcommand{\ee}{\end{equation}}
\newcommand{\ba}{\begin{eqnarray}}
\newcommand{\ea}{\end{eqnarray}}
\newcommand{\bg}{\begin{align}}
\newcommand{\egg}{\end{align}}

\newcommand{\nn}{\nonumber}

\newcommand{\ve}{\epsilon}
\newcommand{\vv}{\mathbf{v}}

\def\bphi{\boldsymbol\phi}

\begin{document}

\title{On the $\phi(1020)\, f_0(980)$ S-wave scattering \\and the Y(2175) resonance}

\author{L.~Alvarez-Ruso, J.~A. Oller and J.~M. Alarc\'on \\
Departamento de F\'{\i}sica. Universidad de Murcia. E-30071,
Murcia, Spain}

\date{June 1, 2009}

\maketitle

\begin{abstract}
We have studied the $\phi(1020)f_0(980)$ S-wave scattering at energies around threshold employing chiral Lagrangians coupled to vector mesons through minimal coupling. The interaction kernel is obtained by considering the $f_0(980)$ as a $K\bar{K}$ bound state. The $Y(2175)$ resonance is generated in this approach by the self-interactions between the $\phi(1020)$ and the $f_0(980)$ resonances. We are able to describe  the $e^+e^-\to \phi(1020)f_0(980)$ recent scattering data to test experimentally our scattering amplitudes, concluding that the $Y(2175)$ resonance has a large $\phi(1020)f_0(980)$ meson-meson component.
\end{abstract}


\section{Introduction}
\label{sec:intro}
\def\theequation{\arabic{section}.\arabic{equation}}
\setcounter{equation}{0}

In recent years, the existence of a new resonance with quantum numbers $J^{PC}=1^{--}$ and mass 
around 2.15~GeV has been revealed by several experiments. In the following we use the notation $Y(2175)$ for this state, although it is also referred as $X(2175)$ or $\phi''$ in the literature. 
The $Y(2175)$ was first observed in the reaction $e^+e^-\to\phi(1020)f_0(980)$  by the BABAR Collaboration~\cite{babar1,babar2}. Its mass and width were determined to be $M_Y=2.175 \pm 0.010 \pm 0.015$~GeV and $\Gamma_Y=0.058 \pm 0.016 \pm 0.020$~GeV~\cite{babar1}. It was then observed in $e^+e^-\to \phi(1020)\eta$ by the same collaboration~\cite{babar3}, though with much less statistical significance, and in  $J/\Psi\to\eta\phi(1020) f_0(980)$ by BES~\cite{bes08} with  $M_Y=2.186\pm 0.010(\mathrm{stat})\pm 0.006(\mathrm{syst})$~GeV and $\Gamma_Y=0.065\pm 0.023(\mathrm{stat})\pm 0.017(\mathrm{syst})$~GeV. The Belle Collaboration~\cite{belle}  has also identified  the $Y(2175)$ in the most precise study so far of the reactions $e^+e^-\to \phi(1020) \pi^+\pi^-$ and $e^+e^-\to\phi(1020)f_0(980)$. The resulting resonance parameters are $M_Y=2.13^{+0.07}_{-0.12}$~GeV and $\Gamma_Y=0.17^{+0.11}_{-0.09}$~GeV. The large errors reflect, among other sources of systematic error, the uncertainties in the determination of the non-resonant background and the possible existence of additional resonances in the vicinity of the $Y(2175)$.  In Ref.\cite{babar2} an  extensive study of the reaction $e^+e^-\to K^+K^-f_0(980)$, with the $f_0(980)$ reconstructed from the $\pi^+\pi^-$ or $\pi^0\pi^0$ signals, is also given. It shows an even more prominent $Y(2175)$ signal than  the $\phi(1020)f_0(980)$ data. The resulting masses then spread  in the range 2.12--2.21~GeV and the width  between  0.045--0.13~GeV, taking into account both central values and the one sigma deviation from them. 

These experimental findings have renewed the theoretical interest in the region of the $Y(2175)$. It has been suggested that this resonance could be a tetraquark state~\cite{wang,hosaka,polosa}. A QCD sum rule calculation taking into account the correlator $(s\bar{s})(s\bar{s})$ between the meson-meson  currents is performed in Ref.~\cite{wang} obtaining $M_Y=2.21 \pm 0.09$~GeV. Both standard and finite energy QCD sum rules are considered in Ref.~\cite{hosaka} with meson-meson and diquark-antidiquark $(ss)(\bar{s}\bar{s})$ currents. The mass value obtained is $2.3\pm 0.4$~GeV.  
Quark models have also been used to address the nature and properties of this resonance. Ref.~\cite{dingyan1} studies the decay modes of the lightest hybrid $s\bar{s}g$ resonance, whose mass was predicted to be in the range 2.1-2.2~GeV~\cite{isgu,closest}, consistently with that of the $Y(2175)$. Its width is estimated to be around 100-150~MeV~\cite{dingyan1}. The identification of the $Y(2175)$ as the quarkonium ($s\bar{s}$) states $2^3D_1$ and $3^3S_1$~\footnote{The spectroscopic notation $n^{2 S+1} L_J$ corresponds to the $n^{th}$ state with spin $S$, orbital momentum $L$ and total angular momentum $J$.}, whose masses have also been predicted to be close to that of the $Y(2175)$~\cite{godfrey}, has been considered in Ref.~\cite{dingyan2}. The $3^3S_1$ assignment is disfavored due to its expected large width, $\Gamma\simeq 0.38$~GeV~\cite{blacky}, while the width of a $2^3D_1$ state is estimated  in the range $0.15$-$0.21$~GeV \cite{dingyan2}. It is argued that the clearly different decay patterns could be used to distinguish between the $2^3D_1$ $\bar{s}s$ and the hybrid $s\bar{s}g$ descriptions \cite{dingyan1,dingyan2}. Instead, Ref.~\cite{polosa} concludes that the diquark-antidiquark picture for the $Y(2175)$ would be characterized  by a prominent $\Lambda\bar{\Lambda}$ decay mode. A Faddeev-type calculation for the $\phi(1020)K\bar{K}$ system is presented in Ref.~\cite{torres} where the interactions between pseudoscalar-pseudoscalar and vector-pseudoscalar mesons are taken from unitarized  Chiral Perturbation Theory (Refs.~\cite{npa} and \cite{roca}, respectively). Remarkably, a peak in the $\phi(1020)K\bar{K}$ strong amplitude  is obtained at the mass of the $Y(2175)$, though the width, around 20~MeV,  is too small. This study indicates that the $Y(2175)$ might have large components corresponding to a resonant $\phi(1020)K\bar{K}$ state.

In the investigation reported here we have studied the S-wave scattering amplitude of the $\phi(1020)f_0(980)$ system  around its threshold. This is feasible because both the $\phi(1020)$ and the $f_0(980)$ are rather narrow resonances. We investigate this process from the theoretical point of view by first deriving the interaction kernel for $\phi(1020)f_0(980)$ and then the scattering amplitude. In this way, by construction, one can distinguish the $\phi(1020)f_0(980)$ $1^{--}$ dynamically generated  bound states or resonances from others pre-existing states or due to genuine three body effects in $\phi(1020)K\bar{K}$ scattering. The needed formalism is elaborated and discussed in section \ref{sec:ff}. We discuss the  appearance of a $\phi(1020)f_0(980)$ resonance with mass and width compatible  with those reported for the $Y(2175)$ \cite{babar1,babar2,belle} in Section~\ref{sec:fb}.  We also reproduce the data on $e^+e^-\to\phi(1020)f_0(980)$ from the same set of references.  Conclusions are given in Section \ref{sec:conclu}. In Appendix~\ref{sec:sup} we discuss the suppression of some of the  diagrams that contribute to the $\phi(1020)f_0(980)$ potential.

 
\section{$\bphi$(1020) $\boldsymbol f_{\boldsymbol0}$(980) scattering}
\label{sec:ff}
\def\theequation{\arabic{section}.\arabic{equation}}
\setcounter{equation}{0}

We first work out the scattering of the  $\phi(1020)$  with a  $K\bar{K}$ state of isospin ($I$) zero, denoted as $|K\bar{K}\ra_0$. Then, we take take advantage of the fact that the  $f_0(980)$ scalar meson is successfully described as a $|K\bar{K}\rangle_0$ bound state~\cite{npa,Janssen:1994wn,Weinstein:1990gu}. Therefore, the $\phi(1020) f_0(980)$ scattering can be determined from the $\phi(1020)K\bar{K}$ one by extracting the residue at the  $f_0(980)$ double pole position that arises from the initial and final  $|K\bar{K}\ra_0$ states.  

We obtain the  different vertices required to determine the $\phi K\bar{K}$ scattering from the lowest order SU(3) chiral Lagrangian \cite{glsu3}
\begin{align}
{\cal L}_2=\frac{f^2}{4}\hbox{Tr}\left(D_\mu U^\dagger D^\mu U+ \chi^\dagger U+\chi U^\dagger \right)\,,
\label{lag2}
\end{align}
with $f$ the pion weak decay constant in the chiral limit, that we approximate to $f_\pi=92.4$~MeV. 
The octet of the lightest pseudoscalar fields are included in $U$ as 
\begin{align}
U&=\exp\left(i\frac{\sqrt{2}\Phi}{f}\right) \,,\nn\\
\Phi&=\frac{1}{\sqrt{2}}\sum_{i=1}^8 \phi_i\lambda_i=\left(
\begin{array}{ccc}\frac{\pi^0}{\sqrt{2}}+\frac{1}{\sqrt{6}}\eta_8 & \pi^+ & K^+\\ 
\pi^- & -\frac{\pi^0}{\sqrt{2}}+\frac{1}{\sqrt{6}}\eta_8 & K^0\\
K^- & \overline{K}^0 & -\frac{2}{\sqrt{6}}\eta_8
 \end{array}\right)\,.
 \label{uu3}
 \end{align}
 The covariant derivative $D_\mu U$ is given by
 \begin{align}
D_\mu U=\partial_\mu U-ir_\mu U+iU \ell_\mu\,,
\label{cov}
\end{align}
with $r_\mu$ and $\ell_\mu$ external right and left fields related to the  vector and axial-vector fields by
\begin{align}
r_\mu&=v_\mu+a_\mu\,,\nn\\
\ell_\mu&=v_\mu-a_\mu\,,
\label{extvec}
\end{align}
respectively. In the following we identify the external vector fields $v_\mu$ with the lightest octet of vector resonances, and the vertices are then determined assuming minimal coupling. This is a generalization of the way in which vector mesons are introduced 
 in vector meson dominance~\cite{saku,bra,prades}. Here we are only interested in the vector fields,
 \be
 r_\mu=g\,v_\mu~,~\ell_\mu=g\,v_\mu\,,
 \label{vec}
 \ee
with $g$ a universal coupling constant. We assume ideal mixing, in terms of which $\phi=-\sqrt{\frac{2}{3}}\omega_8+\frac{1}{\sqrt{3}}\omega_1$
 and $\omega=\frac{1}{\sqrt{3}}\omega_8+\sqrt{\frac{2}{3}}\omega_1$, with $\omega_8$ and $\omega_1$ the $I=0$ octet and singlet vector states, in that order. Whence, 
\be
v_\mu=\left(
\begin{array}{ccc}
\frac{\rho^0}{\sqrt{2}}+\frac{1}{\sqrt{2}}\omega& \rho^+ & {K^*}^+\\
\rho^- & -\frac{\rho^0}{\sqrt{2}}+\frac{1}{\sqrt{2}}\omega & {K^*}^0\\
{K^*}^- & {{\overline{K}^*} }^0 & \phi
\end{array}
\right)\,.
\label{matv}
\ee
  
 As a result, the following Lagrangians involving vector and pseudoscalar mesons arise: 
\begin{align}
{\cal L}_{V^2\Phi^2}&=g^2\hbox{Tr}\left( v_\mu v^\mu \Phi^2-v_\mu \Phi v^\mu \Phi\right)\,,\nn\\
{\cal L}_{V^2\Phi^4}&=-\frac{g^2}{6f^2}\hbox{Tr}\left( v_\mu v^\mu \Phi^4-4 v_\mu
\Phi^3v^\mu \Phi+3 v_\mu \Phi^2 v^\mu \Phi^2\right)\,, \nn\\
{\cal L}_{V\Phi^2}&=-ig\hbox{Tr}\left( v_\mu \Phi \partial^\mu \Phi - v_\mu  \partial^\mu \Phi \Phi \right)\,,\nn\\
{\cal L}_{\Phi^4}&=-\frac{1}{6f^2}\hbox{Tr}\left(\partial_\mu \Phi \partial^\mu \Phi \Phi^2 - \partial_\mu \Phi \Phi \partial^\mu \Phi \Phi -\frac{1}{2} M \Phi^4 \right)\,, 
\label{vvff}
\end{align}
where $M=diag(m_\pi^2,m_\pi^2,2m_K^2-m_\pi^2)$ and $m_\pi$ and $m_K$ the pion and kaon masses. 
 In addition there are vertices of three and four vectors fields which originate from
\begin{align}
{\cal L}_{free}=-\frac{1}{4}\hbox{Tr}\left( F_{\mu \nu}F^{\mu\nu}\right)\,,
\end{align}
with the strength tensor
\begin{align}
F_{\mu\nu}=\partial_\mu v_\nu-\partial_\nu v_\mu-i g[v_\mu,v_\nu]\,.
\end{align}
The resulting Lagrangians involving three and four vector mesons are
\begin{align}
{\cal L}_{V^3}&=i g\hbox{Tr}\left(\partial_\mu v_\nu[v^\mu,v^\nu]\right)\,,\nn\\
{\cal L}_{V^4}&=\frac{1}{2} g^2\hbox{Tr}\left( v_\mu v_\nu[v^\mu,v^\nu]\right)\,.
\label{nonabe}
\end{align} 

\begin{figure}[ht]
\psfrag{k}{$k$}
\psfrag{p}{$p$}
\psfrag{l}{$\ell$}
\psfrag{pi}{$\pi$}
\psfrag{r}{$k-\ell$}
\centerline{\epsfig{file=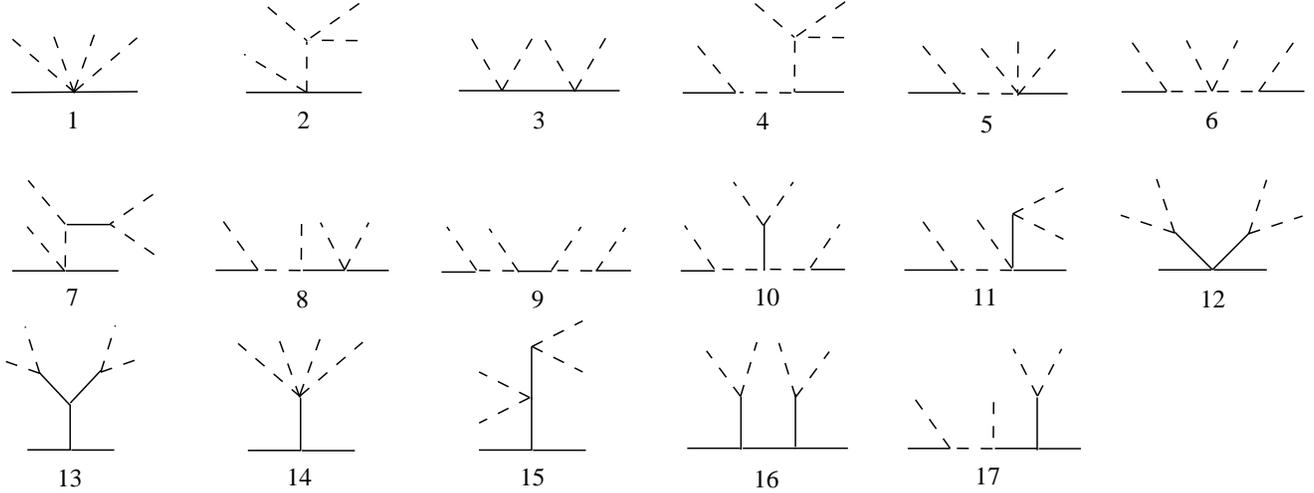,width=1.\textwidth,angle=0}}
\vspace{0.2cm}
\caption[pilf]{\protect \small
Feynman diagrams for $\phi K\bar{K}$ scattering that result from the Lagrangians of Eqs.~(\ref{vvff}, \ref{nonabe}). The dashed lines denote kaons and the solid ones vector mesons ($\phi$ or $\rho_0$). 
\label{fig.dia}}
\end{figure}

The diagrams that contribute to $\phi K\bar{K}\to \phi K\bar{K}$ from the Lagrangians of Eqs.~(\ref{vvff}, \ref{nonabe}) are depicted in Fig.~\ref{fig.dia}.  
Both S- and D- waves contribute to the $\phi(1020) f_0(980)$ scattering in the $1^{--}$ channel but since we are interested in the threshold region around  2~GeV, the D-wave terms can be neglected. They are suppressed by powers of  $|\vp|^{2n}$, where  $|\vp|$ is the three momentum in the center of mass (CM) of the $\phi(1020) f_0(980)$ pair and $n=1,2$ is the number of possible D-wave $\phi(1020)f_0(980)$ states in both  the initial and final scattering states.  
It is also worth stressing that since the $f_0(980)$ is so close to the $K\bar{K}$ threshold, the three-momentum $\vq$ of the kaons in the rest frame of the $f_0(980)$ is small compared to the kaon masses. In this way, a  suppression by powers of  $|\vq|$ and $|\vp|$ can be used to simplify the calculation of the $\phi K\bar{K}$ scattering. On the contrary, the appearance of almost on-shell intermediate mesons enhances some diagrams with respect to the rest. 
Joining both conditions we find that the set of amplitudes represented by the diagram 2 of Fig.~\ref{fig.dia} are dominant because the contributing vertices do not involve any small three-momentum and the intermediate kaon is almost on-shell.  In addition, these diagrams involve an extra large numerical factor because the four-kaon vertex  is around $M_{f_0}^2/f^2\simeq 10^2$, with $M_{f_0}$ the $f_0(980)$ mass. This factor is much larger than the one of a $\phi \phi KK$ vertex,  which scales as $g^2$. Such a vertex appears twice in diagram 3 and once in 2. In spite of the fact that the $\phi$ propagator in diagram 3 is close to its mass shell when one kaon is going in and the other out in each of the vertices, the resulting amplitude is suppressed by more than one order of magnitude with respect to the one from diagram 2 because,  $(M_{f_0}^2/f^2)/g^2$ is large, and also because it  involves less enhanced configurations than the diagram 2.\footnote{An explicit calculation shows that the suppression factor is the inverse of $6 M_{f_0}^2/g^2 f^2\simeq 30$ for $|g|\simeq 5$.} Following similar steps we show in Appendix~\ref{sec:sup} that the rest of diagrams in Fig.~\ref{fig.dia} are suppressed compared with the second one.

\begin{figure}[ht]
\psfrag{k}{$k$}
\psfrag{p}{$p$}
\psfrag{l}{$\ell$}
\psfrag{pi}{$\pi$}
\psfrag{r}{$k-\ell$}
\centerline{\epsfig{file=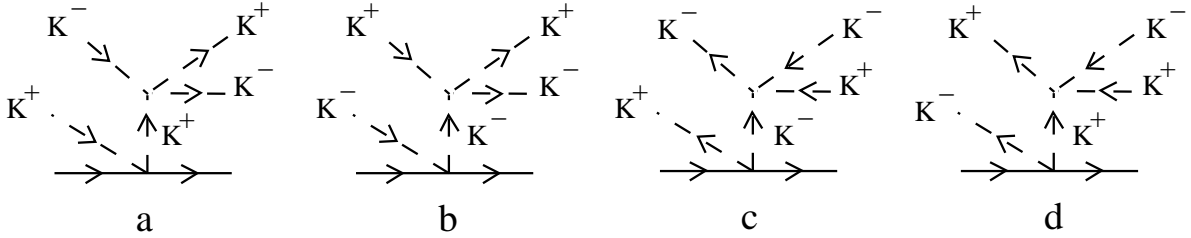,width=.9\textwidth,angle=0}}
\vspace{0.2cm}
\caption[pilf]{\protect \small 
All possible arrangements of kaons in the $\phi(p_1) \, K^+(k_1) \, K^-(k_2) \to \phi(p'_1) \, K^+(k'_1) \, K^-(k'_2)$ scattering from the second diagram of Fig.~\ref{fig.dia} are shown.
\label{fig.pole}}
\end{figure}


Let us proceed with the evaluation of the contribution from diagram 2 of Fig.~\ref{fig.dia} to the $\phi(p_1) \, K^+(k_1) , K^-(k_2) \to \phi(p'_1) \, K^+(k'_1) \, K^-(k'_2)$ scattering represented by the diagrams in Fig.~\ref{fig.pole}. We consider first the $K^+(k) \, K^-(k_2) \to  K^+(k'_1) \, K^-(k'_2)$ amplitude that corresponds to the vertex on the top of Fig.~\ref{fig.pole}(a), with  $k$ the momentum of the intermediate kaon. Using  ${\cal L}_{\Phi^4}$ it can be cast as   
\begin{align}
T^{(a)}_{K^+K^-\to K^+K^-}=-\frac{u_a-2m_K^2}{f^2}-\frac{m_K^2-k^2}{3f^2}\,,
\end{align} 
with $u_a=(k'_1-k_2)^2$. Here the off shell part, which is proportional to the inverse of the kaon propagator, has been explicitly separated; it leads to a contact term in the full amplitude of diagram a. Proceeding analogously with the other three diagrams in Fig.~\ref{fig.pole} and summing all the contributions  it results 
\begin{align}
T_{cc}=&-\frac{8g^2 \ve\cdot \ve'}{3f^2}-\frac{2g^2\ve\cdot \ve'}{f^2}(u_a-2m_K^2)\Bigl[D(Q+k_1)+D(Q-k'_2)\Bigr]\nn\\
&-\frac{2g^2\ve\cdot \ve'}{f^2}(u_b-2m_K^2)\Bigl[D(Q+k_2)+D(Q-k'_1)\Bigr]\,,
\label{tcc}
\end{align}
where $\epsilon$ ($\epsilon'$) is the polarization four-vector of the initial (final) $\phi$ meson, $u_b=(k'_2-k_1)^2$ and $Q=p-p'$.  
The kaon propagator is given by
\begin{align}
D(k)=\frac{1}{m_K^2-k^2-i\varepsilon}\,,
\end{align}
with $\varepsilon\to 0^+$. The subscript $cc$ in $T_{cc}$ indicates that all the kaons are charged. The amplitudes for the $\phi(p_1) \, K^0(k_1) \, \bar{K}^0(k_2) \to \phi(p'_1) \,K^+(k'_1) \,K^-(k'_2)$ and $\phi(p_1) \,K^0(k_1) \,\bar{K}^0(k_2) \to \phi(p'_1) \,K^0(k'_1) \, \bar{K}^0(k'_2)$ reaction channels correspond to diagrams analogous to those in  Fig.~\ref{fig.pole}. Denoting them as $T_{nc}$ and $T_{nn}$, respectively, it reads
\begin{align}
T_{nc}&=\frac{1}{2}T_{cc}~,\nn \\
T_{nn}&=T_{cc}~.
\label{tcn}
\end{align}
Here we have assumed an exact isospin symmetry and used the same values for the masses of charged and neutral kaons. 
Note then that the scattering amplitude for $\phi(p_1) \,K^+(k_1) \,K^-(k_2) \to \phi(p'_1) \,K^0(k'_1) \,\bar{K}^0(k'_2)$, $T_{cn}$, can be obtained from $T_{nc}$ by crossing symmetry and indeed $T_{cn}=T_{nc}$ since the $u$ variables are not altered in the transformation.  

To construct the $I=0$ amplitude we take into account that $|K\bar{K}\ra_{0}$ is 
 \begin{align} 
|K(\vq_1)\bar{K}(\vq_2)\ra_0&=-\frac{1}{\sqrt{2}}|K^+(\vq_1)K^-(\vq_2)+K^0(\vq_1)\bar{K}^0(\vq_2)\ra\,.
\label{isostate}
 \end{align} 
The minus sign appears because we identify $|K^-\ra=-|I=1/2,I_3=-1/2\ra$ to be consistent with the convention adopted in the chiral Lagrangians, Eq.~(\ref{uu3}). Therefore, the resulting $\phi(1020)|K\bar{K}\ra_0\to 
\phi(1020)|K\bar{K}\ra_0$ scattering amplitude from diagram 2 of Fig.~\ref{fig.dia}, $T_{I=0}^{(2)}$, is
\begin{align}
T^{(2)}_{I=0}&=\frac{1}{2}\left\{T_{cc}+T_{nn}+T_{cn}+T_{nc}\right\}=\frac{3}{2}T_{cc}\,.
\label{t2i0}
\end{align}

The contact term in Eq.~(\ref{tcc}) cannot be separated from the one arising from diagram 1 of Fig.~\ref{fig.dia} in a model independent way so we consider this smaller contribution as well. From ${\cal L}_{2V\Phi^4}$ in Eq.~(\ref{vvff}) one has 
\begin{align}
{\cal L}_{\phi^2(K\bar{K})^2}=-\frac{2g^2}{3f^2}\phi_\mu \phi^\mu(K^+K^-+K^0\bar{K}^0)^2\,.
\end{align} 
The resulting contact term, when projected into the $I=0$ channel taking into account Eq.~(\ref{t2i0}) gives 
\begin{align}
 T^{(1)}_{I=0}=-\frac{8g^2}{f^2}\ve\cdot \ve'\,.
 \label{eq.cont}
 \end{align}
Therefore, the resulting $\phi |K\bar{K} \ra_0\to \phi |K\bar{K}\ra_0$ scattering amplitude from the first two diagrams in 
Fig.~\ref{fig.dia} is
\begin{align}
T_{I=0}=\frac{6 g^2}{f^2}\ve\cdot \ve' \Bigl\{ -2 + k_2\cdot k'_1\bigl[D(Q+k_1)+D(Q-k'_2)\bigr]+k_1\cdot k'_2 \bigl[D(Q+k_2)+D(Q-k'_1)\bigr] \Bigr\} \,.
\label{ti0}
\end{align}

\subsection{Extracting the $\boldsymbol f_{\boldsymbol0}$(980) poles}

\begin{figure}[ht]
\psfrag{f0}{{\small $f_0(980)$}}
\centerline{\epsfig{file=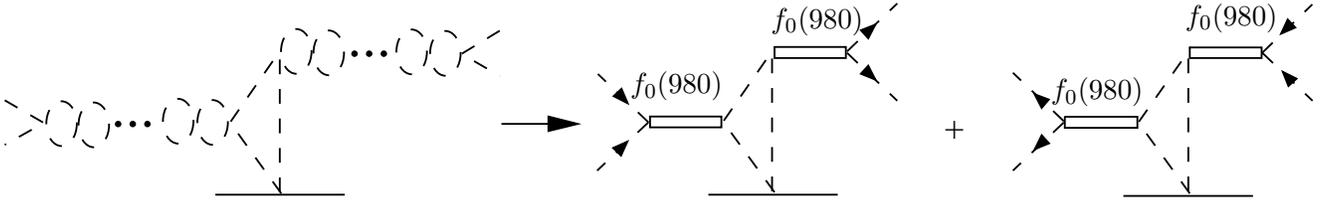,width=1.\textwidth,angle=0}}
\vspace{0.2cm}
\caption[pilf]{\protect \small
The two $f_0(980)$ poles originate because of the $K\bar{K}$ interactions. The kaons and anti-kaons are indicated by the dashed lines.
 \label{fig:itera}}
\end{figure} 

\begin{figure}[ht]
\psfrag{l}{{\small $\ell$}}
\psfrag{l+k}{{ \small $\ell+k$}}
\psfrag{l+k'}{{\small $\ell+k'$}}
\psfrag{l-k1}{{\small $\ell-k$}}
\psfrag{l-k2}{{\small $\ell-k'$}}
\psfrag{Kp}{{\small $K^+$}}
\psfrag{(K0)}{{\small $(K^0)$}}
\psfrag{fi}{{\small $\phi$}}
\psfrag{K}{{\tiny $K$}}
\psfrag{Kb}{{\tiny $\bar{K}$}}
\centerline{\epsfig{file=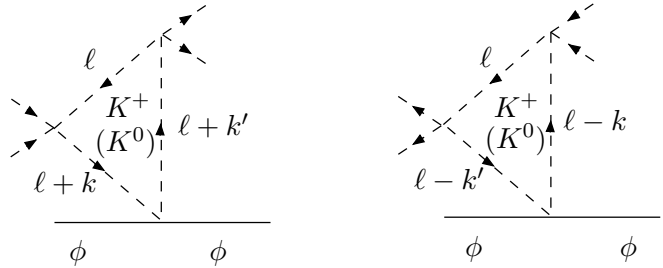,width=.5\textwidth,angle=0}}
\vspace{0.2cm}
\caption[pilf]{\protect \small
Triangular kaon-loop graphs with a $K^+$ or a $K^0$ running in the loop. All the diagrams give the same result. \label{fig:triangle}}
\end{figure}

The $K\bar{K}$ pairs re-scatter giving rise to the diagram shown on the left hand side of Fig.~\ref{fig:itera}. The resulting infinite chain of diagrams contains the poles of the initial and final $f_0(980)$ resonances, as depicted on the right side of the figure. The residue at the $f_0(980)$ double  pole is the $f_0(980)\phi(1020)$ potential $V_{\phi f_0}$. Close to threshold, the two kaons, initial or final, interact predominantly in S-wave. The 
$K\bar{K}$ $I=0$ S-wave amplitude from ${\cal L}_{\Phi^4}$, Eq.~(\ref{vvff}), is~\cite{npa}

\begin{align}
\label{v22.eq}
V_{K \bar{K}}^{\mathrm{full}}(s_{K \bar{K}})=\frac{3 s_{K \bar{K}}}{4f^2}-\frac{1}{4f^2}
\sum_i\left( r_i^2-m_K^2\right) \,,
\end{align}
where $s_{K\bar{K}}$ stands for the invariant mass of the two kaons. The sum runs over all the four kaon states involved in the vertex whose four-momenta are denoted by $r_i$.  The last term is the off-shell part of the amplitude. We use Eq.~(\ref{v22.eq}) in the four-pseudoscalar vertices of the diagrams in Fig.~\ref{fig:triangle}, where $k=k_1+k_2$ and $k'=k'_1+k'_2$ are the total four-momenta of the initial and final $|K\bar{K}\ra_0$ states, respectively. At the $f_0(980)$ double pole $k^2=k'^2=M_{f_0}^2$. A $K^+$ or $K^0$ runs in the loop of the diagrams in Fig.~\ref{fig:triangle}. Taking into account that the $K^+K^- \rightarrow  |K\bar{K}\ra_0$ and $K^0 \bar{K}^0 \rightarrow  |K\bar{K}\ra_0$ vertices are equal to $-\sqrt{1/2} \, V_{K \bar{K}}^{full}$ one gets the same result for the four amplitudes represented in Fig.~\ref{fig:triangle}. 

The off-shell parts comprised in the last term of Eq.~(\ref{v22.eq}), are equal to the inverse of  kaon propagators. In the loops of Fig.~\ref{fig:triangle} they cancel with the kaon propagators giving rise to amplitudes that do not correspond anymore to the dominant triangular kaon-loop but to other topologies so that we disregard them. Therefore, we  obtain  for the sum of the diagrams in Fig.~\ref{fig:triangle} 
\begin{align}
{\cal M}&=4 \epsilon\cdot \epsilon' g^2 V_{K\bar{K}}(k^2)V_{K\bar{K}}(k'^2)\, i \int\frac{d^4\ell}{(2\pi)^4}\frac{1}{\left[\ell^2-m_K^2+i\varepsilon\right]\left[(\ell+k)^2-m_K^2+i\varepsilon\right]\left[(\ell+k')^2-m_K^2+i\varepsilon\right]}
\nn\\
&=-\frac{\epsilon\cdot \epsilon' g^2}{4 \pi^2}  V_{K\bar{K}}(k^2)V_{K\bar{K}}(k'^2) \int_0^1\int_0^1 dz_1 dz_2 \frac{\theta(1-z_1-z_2)}{k^2 z_1(1-z_1)+ k'^2 z_2(1-z_2)-2 k\cdot k' z_1 z_2-m_K^2+i\varepsilon}\,,
\end{align}
where $V_{K\bar{K}}$ is the on-shell part of Eq.~(\ref{v22.eq}) given by its first term. Next, we perform the  following change of integration variables 
\begin{align}
x=&\frac{1}{2}(z_1+z_2)~,\nn\\
y=&z_1-z_2~.
\end{align}
After performing the integration on $y$ we have
\begin{align}
{\cal M}=&\epsilon\cdot\epsilon' g^2V_{K\bar{K}}(k^2)V_{K\bar{K}}(k'^2)\frac{1}{\pi^2Q^2}\int_0^{1/2} dx
\frac{\log\left(1-2x/c\right)-\log\left(1+2x/c\right)}{c}~,
\label{mt}
\end{align}
with
\begin{align}
c=&\sqrt{-\frac{16 k^2}{Q^2}\left(1-\frac{Q^2}{4k^2}\right)\left[(x-x_0)^2+a^2\right]}\,,\nn\\
x_0=&\frac{1}{4 \left(1-\frac{Q^2}{4k^2}\right)}~,\nn\\
a^2=&\frac{1}{16 \left(1-\frac{Q^2}{4k^2}\right)^2}\left[\frac{4m_K^2}{k^2}\left(1-\frac{Q^2}{4k^2}\right)-1\right]\,.
\label{cg2}
\end{align}
Inside the integral we take $k^2=k'^2$, which is correct at the $f_0(980)$ double pole.
Next, we have to project into the S-wave state of the $\phi(1020)f_0(980)$ system, which amounts to integrating over $\cos\rho\in[-1,1]$, with $\rho$ the relative angle between $\vp$ and $\vp'$ in the CM frame. In terms of it $Q^2=-2\vp^2(1-\cos\rho)$. The leading non-relativistic contribution for $\ve(\vp,s)\cdot \ve'(-\vp,s')=-\delta_{ss'}+{\cal O}(\vv^2)$, with $\vv=\vp/W$ and $W$ denotes the total CM energy of the $\phi(1020) f_0(980)$ pair. Since $\vv$ is small we just keep the first term and replace $\ve\cdot \ve'\to -1$  in Eq.~(\ref{mt}) and in the tree-level contact term of Eq.~(\ref{ti0}) that we add to the former, obtaining the S-wave amplitude 
\begin{align}
{\cal M}^S_{I=0}=&\frac{12 g^2}{f^2}-V_{K\bar{K}}(s_1)V_{K\bar{K}}(s'_1)\frac{g^2}{2\pi^2}\int_{-1}^{+1}\frac{d\cos\rho}{Q^2}
\int_0^{1/2} dx
\frac{\log\left(1-2x/c\right)-\log\left(1+2x/c\right)}{c} \,.
\label{eq.ms}
\end{align}
One should bear in mind that some of the discarded contributions to the triangle loops from the off-shell parts of \eqref{v22.eq} lead to contact terms that would just renormalize the first term of the previous equation.  

The next step is to resum the re-scattering chain for each of the $(K\bar{K})_0$ pairs, as represented in the left diagram of Fig.~\ref{fig:itera}. This can be done by multiplying ${\cal M}^S$ by the factor \cite{ddecays} 
\begin{align}
&\frac{1}{D(k^2)D(k'^2)}\,,
\label{resc2}
\end{align}
with 
\begin{align}
D(k^2)=1+V_{K\bar{K}}(k^2) G_2(k^2)~.
\label{resc22}
\end{align}
The function $G_2$ represents the unitary loop of two kaons.  
In Ref.~\cite{npa} it was established that the $f_0(980)$ is predominantly a $|K\bar{K}\ra_0$ S-wave bound state that slightly modifies its mass and acquires a  narrow width due to the coupling to pions. 
Then, in order to reproduce the $f_0(980)$ pole properties due to its coupling to kaons, one can consider single channel kaon scattering and write~\cite{npa} 
\begin{align}
T_{K\bar{K}}(k^2)&=\frac{V_{K\bar{K}}(k^2)}{1+V_{K\bar{K}}(k^2)G_{2}(k^2)}=\frac{V_{K\bar{K}}(k^2)}{D(k^2)}\,.
\label{t22}
\end{align} 
This equation can be interpreted as the evolution of a $|K\bar{K}\ra_0$ pair produced by the potential $V_{K\bar{K}}$ that undergoes re-scattering  as determined by the factor $[(1+V_{K\bar{K}}(k^2)G_2(k^2)]^{-1}=1-V_{K\bar{K}}G_2+V_{K\bar{K}} G_2 V_{K\bar{K}} G_2 +\ldots$. For our present problem on the $\phi(1020) f_0(980)$ scattering two $|K\bar{K}\ra_0$ pairs re-scatter by initial and final state interactions. Analogously, from Eq.~(\ref{eq.ms}) one has
\begin{align}
 M^S &=\frac{{\cal M}^S}{D(k^2)D(k'^2)}\nn\\
&=\left[\frac{12 g^2/f^2}{V_{K\bar{K}}(k^2) V_{K\bar{K}}(k'^2)}-\frac{g^2}{2\pi^2}\int_{-1}^{+1}\frac{d\cos\rho}{Q^2} \int_0^{1/2} dx
\frac{\log\left(1-2x/c\right)-\log\left(1+2x/c\right)}{c}\right] \nn\\
& \times T_{K\bar{K}}(k^2)T_{K\bar{K}}(k'^2)\,.
\label{eq.mms}
\end{align}
It is worth stressing here that this equation  can be interpreted as a purely phenomenological one corresponding to the topology of the diagrams 1 and 2 of Fig.~\ref{fig.dia}. It is parameterized in terms of the $K\bar{K}$ $I=0$ S-wave amplitude $T_{K\bar{K}}$. The first term corresponds to a general contact interaction at threshold.\footnote{In our fits to data (see next section) we have allowed two different values of $g^2$, one for the contact term and another for the kaon pole terms in Eq.~(\ref{eq.mms}). However, we have not found any significant difference in our conclusions so that we skip any further comment on this issue.}

The scattering amplitude $T_{K\bar{K}}(k^2)$ has a pole below the $K\bar{K}$ threshold due to the $f_0(980)$ bound state, which implies that
\begin{align}
\lim_{k^2\to M_{f_0}^2} (M_{f_0}^2-k^2)T_{K\bar{K}}(k^2)= \gamma_{K\bar{K}}^2\,.
\label{resi}
\end{align}
Then, at $k^2,$ $k'^2\to M_{f_0}^2$\,, 
\begin{align}
V_{\phi f_0}&=\frac{1}{\gamma_{K\bar{K}}^2}\lim_{k^2, k'^2\to M_{f_0}^2}(k^2-M_{f_0}^2)(k'^2-M_{f_0}^2)M^S 
\nn\\
&=
\left[\frac{12 g^2}{f^2 V_{K\bar{K}}(M_{f_0}^2)^2}-\frac{g^2}{2\pi^2}\int_{-1}^{+1}\frac{d\cos\rho}{Q^2}
\int_0^{1/2} dx
\frac{\log\left(1-2x/c\right)-\log\left(1+2x/c\right)}{c}\right]\gamma_{K\bar{K}}^2~.
\label{vff}
\end{align} 
The coupling of the $f_0(980)$ to  $|K\bar{K}\ra_0$, $\gamma_{K\bar{K}}$, has the value $\gamma_{K\bar{K}}\simeq 4$~GeV \cite{nd,mixing}. The $1/\gamma_{K\bar{K}}^2$ factor appears because $M^S$ contains two extra couplings $f_0(980)\to |K\bar{K}\ra_0$ that should be removed when isolating the $f_0(980)$ resonances.    

Finally,  the  $\phi(1020)f_0(980)$ S-wave scattering amplitude is obtained by an  expression analogous to Eq.~\eqref{t22},
\begin{align}
T_{\phi f_0}=\frac{V_{\phi f_0}}{1+V_{\phi f_0} G_{\phi f_0}}~.
\label{tff}
\end{align}
For a general derivation of this equation based on the N/D method see Refs.\cite{nd,kn}. 
Here, $G_{\phi f_0}$ is the unitary loop function of a $\phi(1020)$ and a $f_0(980)$ resonances and is given by \cite{nd,prd}
\begin{align}
G_{\phi f_0}(s)&=\frac{1}{(4\pi)^2}\biggl\{
a_1+\log\frac{M_{f_0}^2}{\mu^2}-\frac{M_\phi^2-M_{f_0}^2+s}{2s}\log\frac{M_{f_0}^2}{M_\phi^2}
+\frac{|\vp|}{\sqrt{s}}\biggl[\log(s-\Delta+2\sqrt{s}|\vp|)\nn\\
&+\log(s+\Delta+2\sqrt{s}|\vp|)
-\log(-s+\Delta+2\sqrt{s}|\vp|)-\log(-s-\Delta+2\sqrt{s}|\vp|)
\biggr]\biggr\}~,
\label{gff}
\end{align}
with $\Delta=M_{\phi}^2-M_{f_0}^2$. While the renormalization scale $\mu$ is fixed to value of the $\rho$ meson mass, $\mu=770$~MeV, the subtraction constant $a_1$ has to be fitted to data~\cite{nd}.

%
\section{ $\bphi$(1020) $\boldsymbol f_{\boldsymbol0}$(980) resonant states}
\label{sec:fb}
\def\theequation{\arabic{section}.\arabic{equation}}
\setcounter{equation}{0}

\begin{figure}[ht]
\psfrag{nb}{$|T_{\phi f_0}|^2$}
\psfrag{GeV}{$\begin{array}{c}
\\
\sqrt{s} \hbox{[GeV]}
\end{array}$}
\centerline{\epsfig{file=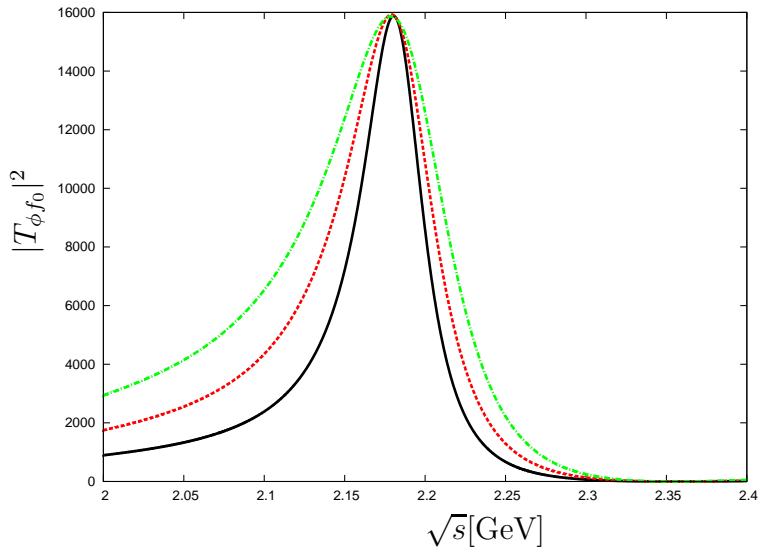,width=.4\textwidth,angle=-90}}
\vspace{0.2cm}
\caption[pilf]{\protect \small
$|T_{\phi f_0}|^2$ with the peak at $2.18$~GeV as a function of the $\phi f_0$ invariant mass. The solid, dashed and dot-dashed lines are  for $\sqrt{-g^2}=5,$~6,~7 and $a_1=-7.1$~,$-5.2$, $-4.1$ respectively.
\label{fig.peaks}}
\end{figure}

The potential $V_{\phi f_0}$, Eq.~(\ref{vff}), depends on $g^2$ mainly through the vertex at the bottom of the diagrams of Fig.~\ref{fig.pole} that corresponds to $\phi(1020) K$ scattering. In the present problem on the $\phi(1020)f_0(980)$ scattering around its threshold we are also close to the $\phi(1020)K$ threshold itself. On the other hand, the $K_1(1400)$ resonance is only 100~MeV 
below it. Therefore, it is quite reasonable to expect that the $\phi(1020)K$ scattering is dominated by this resonance which implies 
that $g^2<0$ because, for a bare pole,
\begin{align}
g^2\sim \frac{\gamma_{K_1\phi K}^2}{M_{K_1}^2-(M_\phi+m_K)^2}<0~.
\end{align} 
 In this way, $g^2$ is interpreted as a parameter that mimics the $\phi(1020)K$ scattering amplitude in the energy region of the $\phi(1020)f_0(980)$ scattering close to threshold. On the other hand,  we restrict $g^2$ to be real so that $V_{\phi f_0}$ is also real above the $\phi(1020)f_0(980)$ threshold and the resulting S-wave $T_{\phi f_0}$ amplitude, Eq.~(\ref{tff}), fulfils unitarity.\footnote{We have checked that our fits to data are stable if we allow $g^2$ to become complex.}  
With $g^2<0$, $V_{\phi f_0}$ is  positive (attractive)  around the $\phi(1020) f_0(980)$ threshold. In this situation $|T_{\phi f_0}|^2$ has resonant peaks with mass and width compatible with those measured for the  $Y(2175)$~\cite{babar2,bes08,belle}. The $Y(2175)$ mass and width values extracted by BABAR~\cite{babar2} and BES~\cite{bes08} are compatible between each other. In the following we take their average 
\begin{align}
M_Y&=2.180\pm 0.008~\hbox{GeV}\,,\nn\\
\Gamma_Y&=0.060\pm 0.014~\hbox{GeV}
\label{yvalues}
\end{align}
as reference values. In Fig.~\ref{fig.peaks} we show $|T_{\phi f_0}|^2$ for $(\sqrt{-g^2},a_1)=(5,-7.1),$ $(6,-5.2)$ and $(7,-4.1)$ with $M_{f_0}=0.98$~GeV in  all the curves. The peak is located at $2.18$~GeV as in Eq.~(\ref{yvalues}) and the width increases with $\sqrt{-g^2}$, taking the values of $48$, $72$ and $100$~MeV for $\sqrt{-g^2}=5$,~6 and 7, in that order. Although the width increases, the size at the peak remains constant because the former is proportional to $g^2$ so that the ratio $g^2/\Gamma_Y$, which fixes the amplitude at the maximum, is roughly independent of the value of $g^2$ used for a fixed peak position. 

To sharpen our conclusions we now compare directly with the $e^+e^-\to \phi f_0(980)$ data~\cite{babar1,babar2,belle}. The $\phi(1020)f_0(980)$ strong scattering  amplitude, Eq.~(\ref{tff}), is employed to correct by final state interactions (FSI) a given production process for $e^+e^-\to \phi f_0(980)$. This is achieved \cite{ddecays} by multiplying the production amplitude by 
\begin{align}
\frac{1}{1+G_{\phi f_0}V_{\phi f_0}}\,,
\label{fsi}
\end{align}
in the same manner as already done in Eq.~(\ref{resc2}) for the $K\bar{K}$ re-scattering.   
We take as the non-resonant production cross section $\sigma_{NR}(s)$ the one fitted in Fig.~6(b) of Ref.~\cite{belle}. Therefore, after FSI, 
\begin{align}
\sigma_{NR}(s)\to \sigma_R(s)=\frac{\sigma_{NR}(s)}{\left|1+V_{\phi f_0}{G}_{\phi f_0}\right|^2}~.
\label{sigr}
\end{align} 
In order to take into account the mass distribution of the $f_0(980)$ resonance, the previous result is convoluted with the $f_0(980)$ mass distribution $P(W_{f0})$ 
\begin{align}
\la\sigma_R(s)\ra ={\cal N} \sigma_{NR}(s) \int dW_{f_0}\frac{P(W_{f_0})}{\left|1+V_{\phi f_0}{G}_{\phi f_0}(s,W_{f_0})\right|^2}\,,
\label{srave}
\end{align}
For practical purposes we take $P(M_{f_0})$ as a Lorentzian distribution centered at $M_{f_0}=0.98$ or 0.99~GeV with a width $\Gamma_{f_0}=$50~MeV~\cite{alba},
\begin{align}
P(W_{f_0})&=\frac{1}{2\pi}\frac{\Gamma_{f_0}}{\displaystyle (W_{f_0}-M_{f_0})^2+\frac{\Gamma_{f_0}^2}{4}}\,.
\end{align}
The normalization constant ${\cal N}$ is included in Eq.~\eqref{srave} to account for the fact that 
the $\sigma_{NR}(s)$ of Ref.\cite{belle} is extracted assuming an specific shape and strength for the 
resonant signal. 

\begin{figure}[h!]
\psfrag{nb}{$\sigma$ [nb]}
\psfrag{GeV}{$\begin{array}{c}
\\
\sqrt{s} \,\hbox{[GeV]}
\end{array}$}
\psfrag{sig para fi(1020)f0(980) g2 negativa ahora}{$\sigma(e^+e^-\to \phi(1020)f_0(980))$}
\centerline{\epsfig{file=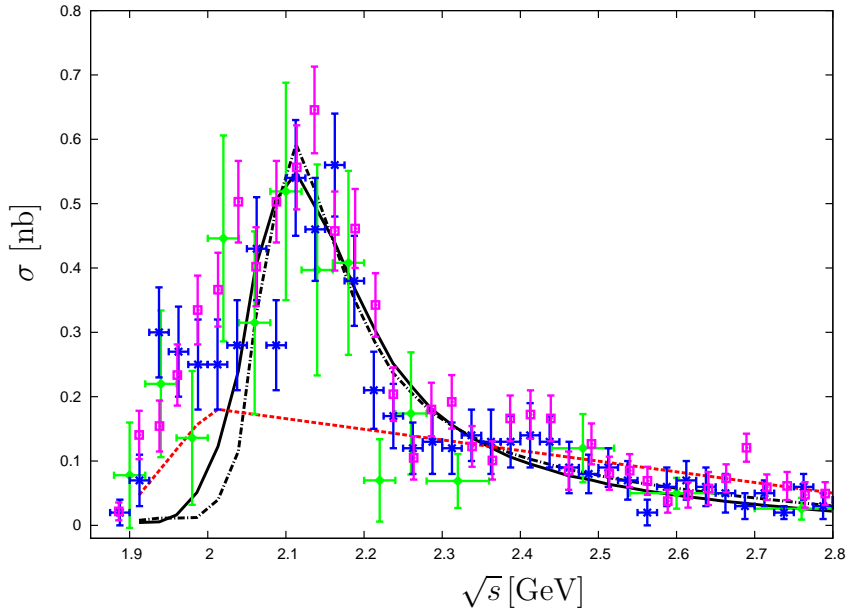,width=.45\textwidth,angle=-90}}
\vspace{0.2cm}
\caption[pilf]{\protect \small
Cross-section for $e^+e^-\to \phi(1020)f_0(980)$. The experimental data are from Ref.~\cite{babar2} (diamonds and crosses) and Ref.~\cite{belle} (empty boxes). The solid and dash-dotted lines correspond to the first and second fits of Table~\ref{tab.gneg.fit}. The dashed line shows ${\cal N}\sigma_{NR}(s)$ for the first fit.
\label{fig.gneg.fits}}
\end{figure}

\begin{table}[h!]
\begin{center}
\begin{tabular}{|l|l|l|l|l|}
\hline
  $\sqrt{-g^2}$  & $a_1$ & ${\cal N}$ & $\chi^2_{d.o.f.}$ & $M_{f_0}$  \\
\hline
$7.33\pm 0.30$ & $-2.41\pm 0.14$ & $0.79\pm 0.06$ & $88/(46-3)$  & 0.98~GeV (fixed)\\
\hline
$3.94\pm 0.18$ & $-2.84\pm 0.18$ & $0.52\pm 0.05$ & $108/(46-3)$  &  0.99~GeV (fixed)\\
\hline
\end{tabular}
\caption{Fits to the data from BABAR \cite{babar2} and Belle \cite{belle} on $e^+e^-\to \phi(1020)f_0(980)$. The first fit uses $M_{f_0}=0.98~$GeV and the second one $M_{f_0}=0.99$~GeV. 
\label{tab.gneg.fit}}
\end{center}
\end{table}
We have performed fits using the data points around the $Y(2175)$ peak, for $\sqrt{s}\in  [2,2.6]$~GeV, taking into account the bin size.  The best-fit parameters for $M_{f_0}=0.98$ and $0.99$~GeV are given in Table~\ref{tab.gneg.fit}.  
The results of these fits are the solid and dot-dashed lines in Fig.~\ref{fig.gneg.fits}, where the points used to draw the curves are separated in energy according to the bin size of the experimental points of Refs.\cite{belle,babar2} for the data set from the $\phi(1020)\pi^+\pi^-$ final state. The rest of  the points (diamonds) are obtained from the  $\phi(1020)\pi^0\pi^0$ final state~\cite{babar2}.
Notice that the data from Ref.~\cite{belle} are slightly more precise than those from Ref.~\cite{babar1,babar2}. The fitted parameters do not depend on the precise value of the upper energy limit. We have used $\sqrt{s}=2.6$~GeV as a large enough value to cover the energy region where our approach is valid, namely, near the $\phi(1020)f_0(980)$ threshold. The suppression of our results for $\sqrt{s}\lesssim 2$~GeV  in Fig.~\ref{fig.gneg.fits} is not due to a negative interference of $T_{\phi f_0}$ with the non-resonant contribution. Instead, it is due to fact that  at the $\phi(1020)f_0(980)$ threshold, the $V_{\phi f_0}$ potential of  Eq.~(\ref{vff}) is large because of the $1/Q^2$ factor.
This threshold effect is very sensitive to the procedure to disentangle the $f_0(980)$ resonant signal. Here, we have taken for it a precise mass value given by the $f_0(980)$ pole position. However, experimentally it is obtained from  the $e^+e^-\to\phi(1020)\pi\pi$ data by integrating the two pion invariant mass distribution within an energy  region around the $f_0(980)$ signal, typically for $0.85$~GeV$\leq \sqrt{s_{\pi\pi}}\leq 1.1$~GeV \cite{babar1}, with $\sqrt{s_{\pi\pi}}$ the two pion invariant mass.

\begin{figure}[ht]
\psfrag{nb}{$|V_{\phi f_0}|$}
\psfrag{GeV}{$\begin{array}{c}
\\
\sqrt{s} \,\hbox{[GeV]}
\end{array}$}
\centerline{\epsfig{file=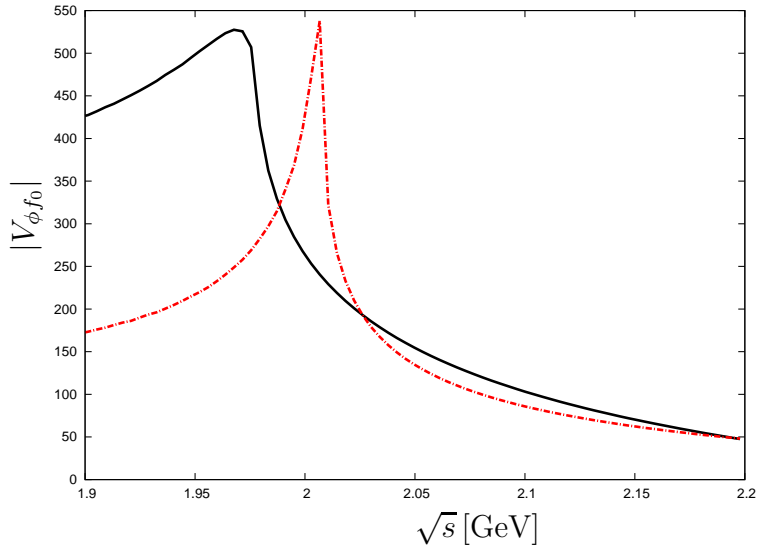,width=.4\textwidth,angle=-90}}
\vspace{0.2cm}
\caption[pilf]{\protect \small
$|V_{\phi f_0}|$ for the two sets in Table~\ref{tab.gneg.fit}. The solid and dot-dashed lines correspond to the first and second fits, respectively.
\label{fig:pot}}
\end{figure}

  In Fig.~\ref{fig:pot},  $|V_{\phi f_0}|$ is shown for the two sets of parameters given in Table~\ref{tab.gneg.fit}. The solid line is for $M_{f_0}=0.98$~GeV and the dot-dashed one for $M_{f_0}=0.99$~GeV. Both have a similar peak value, though to accomplish this $|g^2|$ is smaller by around a factor 3 for the first fit in Table~\ref{tab.gneg.fit} compared to the second. The reason is again related to the factor $1/Q^2$ in $V_{\phi f_0}$, Eq.~\eqref{vff}. Indeed, the integration in $x$ is logarithmically divergent for those $Q^2>0$ values (below the $\phi(1020)f_0(980)$ threshold) which are large enough [around $ M_{f_0}^2(4 - M_{f_0}^2/m_K^2)$] to make $a^2$ in Eq.~(\ref{cg2}) vanish. However, the logarithmic divergence in $x$ disappears after the integration in $\cos \theta$ is performed. The onset of this behavior gives rise to the maximum of $V_{\phi f_0}$ below threshold, as can be seen in Fig.~\ref{fig:pot}.  There is an exception for which the logarithmic divergence in $x$ remains; this occurs exactly at the $\phi(1020)f_0(980)$ threshold and only for $M_{f_0}=2m_K$. In this case $a^2=0$ for all $\cos\theta$ and the final result after the two integrations is logarithmically divergent at threshold. This is the reason why, for a fixed value of $g^2$, as $M_{f_0}$ approaches $2m_K$ the potential becomes larger with a narrower peak structure. This is the limit that corresponds exactly to the suppression mechanisms used to establish that the diagram 2 of Fig.~\ref{fig.dia} is the dominant one. The appearance of the $Y(2175)$ peak within our approach is driven by the large value of $V_{\phi f_0}$ at threshold and its rather fast decrease in energy for $\sqrt{s}$ somewhat above the $\phi(1020)f_0(980)$ threshold. 
 
It is interesting to mention that while $\sqrt{-g^2}$ in Table~\ref{tab.gneg.fit} is in the range of values used in Fig.~\ref{fig.peaks},  the $a_1$ values in the table are smaller in modulus  by around a factor 2--3 compared to those used in Fig.~\ref{fig.peaks}, to obtain $M_Y=2.18$~GeV as in Eq.~\eqref{yvalues}.  This implies that $|T_{\phi f_0}|^2$ from the fits to data has a peak at smaller energies (around 2.09~GeV) and wider, with a width of around 150~MeV. This is in line with the findings of the Belle Collaboration \cite{belle} discussed in the Introduction. In all cases $a_1$ is negative, as it should be for a dynamically generated resonance. In this situation the potential should be attractive so that $1/V_{\phi f_0}$ can cancel with $a_1$ in Eq.~(\ref{tff}). On the other hand, for the $a_1$ values in Table~\ref{tab.gneg.fit}, the resulting unitary $\chi PT$ scale, $\Lambda=(4\pi f)/\sqrt{|a_1|}\simeq 0.75$~GeV,  preserves a natural size around $M_\rho$. However,  since the $|a_1|$ values in Fig.~\ref{fig.peaks} are significantly larger, the interpretation of these peaks as fully dynamically generated states is more arguable because in this case the unitarity scale is just around $0.5$~GeV. Nonetheless, even in these cases one can still conclude that these peaks have a large $\phi(1020)f_0(980)$ re-scattering component. On the other hand, for a resonance mass $M_Y=2.09$~GeV one has $1/p=0.65$~fm and for $M_Y=2.18$~GeV, $1/p=0.45$~fm. This indicates that although the $Y(2175)$ had large a $\phi(1020)f_0(980)$ meson-meson components, as our results point out, it is a rather compact object.

\section{Conclusions}
\label{sec:conclu}

We have studied the $\phi(1020) f_0(980)$ S-wave dynamics in the threshold region. First, the $\phi K\bar{K}$ scattering amplitude at tree level has been determined from the chiral Lagrangians using minimal coupling. The re-scattering of the two kaons in an $I=0$, S-wave state gives rise to the $f_0(980)$ as a bound state. The residue at the $f_0(980)$ double pole in the initial and final states is used to determine the interaction potential between the resonances $\phi(1020)$ and $f_0(980)$ without introducing new extra free parameters. Afterwards, the $\phi(1020)f_0(980)$ S-wave scattering amplitude is determined by resuming the unitarity loops or right hand cut. Resonant peaks with mass and width in agreement with those of the $Y(2175)$ are naturally obtained within the approach. In addition, we are able to describe the $e^+ e^-\to\phi(1020)f_0(980)$ experimental data \cite{babar1,babar2,belle} in terms of the resulting $\phi(1020)f_0(980)$ S-wave amplitude using natural values of the coupling $g^2$ and the subtraction constant $a_1$. The negative value of $g^2$ is reasonable, viewed as a parameter that accounts for the $\phi K$ scattering above the $K_1(1400)$ resonance. The negative values of $a_1$ is characteristic of dynamically generated resonances. Nonetheless, the $a_1$ values required to obtain the $Y(2175)$ resonance at the nominal mass of 2.18~GeV \cite{babar2,bes08} are larger in modulus than those obtained in our direct fits to $e^+e^-\to\phi(1020)f_0(980)$ data. The latter values fit better for interpreting the $Y(2175)$ as mainly a $\phi(1020)f_0(980)$ dynamically generated resonance while the former ones tend to indicate some extra (preexisting) contribution. Taking into account both possibilities, our results suggest that the $Y(2185)$ is a resonant with at least a large $\phi(1020)f_0(980)$ component.

\section*{Acknowledgements}
 We would like to thank  L.~Roca and A. Martinez for useful discussions, and E. Sodolov for his assistance with the BABAR data.  This work is partially funded by the grant MEC  FPA2007-6277 and by 
the BMBF grant 06BN411,  EU-Research Infrastructure
Integrating Activity
 ``Study of Strongly Interacting Matter" (HadronPhysics2, grant n. 227431)
under the Seventh Framework Program of EU 
and HGF grant VH-VI-231 (Virtual Institute ``Spin and strong QCD''). L.~Alvarez-Ruso would like to thank the Fundaci\'on S\'eneca (Murcia) for funding his stay at the Physics Department of the University of Murcia, and the latter for its warm hospitality.

\section*{Appendices}

\appendix{}

\section{Suppression of diagrams in Fig.~\ref{fig.dia}}
\label{sec:sup}
\def\theequation{\Alph{section}.\arabic{equation}}
\setcounter{equation}{0}

Let us now consider the relative size of diagrams 4--17 in Fig.~\ref{fig.dia} compared to diagram 2. Diagram 3 was already discussed at the end of section \ref{sec:ff}. 

{\bf Diagram 4}

The enhanced configurations are those in which the kaon line on the left correspond to an outgoing particle. 
In this way the leftmost intermediate kaon propagator is almost on-shell, taking the value
 \begin{align}
 \frac{1}{(p-k'_{1,2})^2-m_K^2}~.
 \end{align}
 At threshold it is given by  $1/a$ with 
 \begin{align}
a=2m_K(M_\phi-2m_K)\equiv 2m_K\delta\,.
 \label{adef}
 \end{align}
Numerically $\sqrt{a}\simeq 170$~MeV. Like in diagram 2, the vertical kaon propagator is nearly on-shell and of value $1/a'$ with 
\begin{align}
a'=(k_1+k_2-k'_{2,1})^2-m_K^2=(p'-p+k'_{1,2})^2-m_K^2\,.
\label{ap1}
\end{align}
While $a$ is not large because of the proximity of the $\phi$ mass to the $K\bar{K}$ threshold, $a'$ is proportional to the small kaon three-momenta. The initial and final $|K\bar{K}\ra_0$ states are not in their CM. The velocities of the boosts that take these states to their CM frames are $\vv=-\vp/\sqrt{s}$ and $\vv'=-\vp'/\sqrt{s}$, respectively. These velocities are small because we are close to threshold so that 
it is a good approximation to write
\begin{align}
\vk_{1,2}=\pm\vq+m_K\vv+{\cal O}(|\vv|^3)=\pm \vq-\frac{2m_K \sqrt{s}}{s+M_{f_0}^2-M_\phi^2}\vp+{\cal O}(|\vv|^3)\simeq \pm \vq -\frac{1}{2}\vp\,.
\label{boost1}
\end{align}
Similarly,
\begin{align}
\vk'_{1,2}=\pm\vq'+m_K\vv'+{\cal O}(|\vv|^3)=\pm \vq'-\frac{2m_K \sqrt{s}}{s+M_{f_0}^2-M_\phi^2}\vp'+{\cal O}(|\vv|^3)\simeq \pm \vq' -\frac{1}{2}\vp'\,.
\label{boost2}
\end{align}
In the last two equations $\vq$ and $\vq'$ are the CM three-momentum of a kaon in the initial and final $|K\bar{K}\ra_0$ states, respectively. The last equality in these equations follows because $2m_K \sqrt{s}/(s+M_{f_0}^2-M_\phi^2)\simeq 1/2$. In this way we can rewrite  Eq.~(\ref{ap1}) as
\begin{align}
a'=(p'-p+k'_{1,2})^2-m_K^2=Q^2-2Q k'_{1,2}&\simeq -\vp^2(1-\cos\rho)\pm 2\mathbf{Q}\vq'\,,
\label{ap2}
\end{align}
which is zero at threshold. On the other hand, the vertices involving the coupling of the external vector resonances to two-kaons are proportional to small three-momenta. As a result, this diagram is of order
\begin{align}
\frac{g^2}{f^2}\frac{m_K^2}{a'}\frac{|\vk|^2}{a}\,,
\end{align}
with $|\vk|$ representing the modulus of any small external three-momentum.
Since $|\vk|^2/a={\cal O}(1)$, this diagram seems to be of the same order as diagram 2. However, there is an extra suppression coming from the angular projection into S-wave. The angular dependence is dominated by the ratio $|\vk|^2/a'$, since $a$ has a finite angular independent part [Eq.~\eqref{adef}]. From the vertices with one vector resonance one gets a factor
 \begin{align}
 \epsilon(p)\cdot k'_1 \epsilon(p')\cdot (k'_1-p)~.
 \label{eps.4}
 \end{align}
If the spin direction is given by the unitary vector $\hat{\mathbf{n}}$, such that $\hat{\mathbf{n}}\cdot\vp=0$ we can write 
 \begin{align}
 \epsilon(p)&=(0,\hat{\mathbf{n}})~,\nn\\
 \epsilon(p')&=(\vp'\cdot \hat{\mathbf{n}}/p'_0,\hat{\mathbf{n}})
 +{\cal O}(\vv^2)~.
 \label{epsvec}
 \end{align} 
 The following angular structures result from Eq.~\eqref{eps.4} 
\begin{align}
&(\vp'\cdot \hat{\mathbf{n}})^2 ~,\nn\\
&(\vp'\cdot \hat{\mathbf{n}})  (\hat{\mathbf{n}}\cdot \vq')~,\nn\\
&( \hat{\mathbf{n}}\cdot \vq')^2~.
\end{align} 
In the energy region where $|\vp|\lesssim \sqrt{a}$, diagram 4 is suppressed compared to diagram 2 because of the ratio $|\vk|^2/a$. On the other hand, for $|\vp|\gtrsim \sqrt{a}$, $a'$ is dominated by $\vp^2(1-\cos\rho)$ because $\vp^2\gg \vq^2 \approx M_{f_0}^2/4-m_K^2$ for typical energies just slightly above threshold. Thus, we can neglect its angular dependence on ${\hat{\vq}}'$ in good approximation. In this case, the second angular structure in the previous equation vanishes because of the integration in ${\hat{\vq}}'$. The last structure is suppressed by a factor $\kappa^2$, with  $\kappa=|\vq|/|\vp|$. Regarding the dominant structure in the first line, when divided by $a'$, one has
\begin{align}
\int_{-1}^{+1} d\cos\rho
\frac{\vp^2 \sin^2\rho\cos^2\phi}{\vp^2(1-\cos\rho)}~,
\label{intrho}
\end{align}
where we have written $(\vp'\cdot\hat{\mathbf{n}})^2=\vp^2\sin^2\rho \cos^2\phi$, since $\hat{\mathbf{n}}$ is perpendicular to $\vp$. Equation~\eqref{intrho} is finite and equal to $2\cos^2\phi$. For diagram 2 the angular integration on $\hat{\vp}'$ is dominated by  $1/a'$,
\begin{align}
\int_{-1}^{+1} d\cos\rho \frac{1}{\vp^2(1-\cos\rho)}~,
\end{align}
which is divergent. Keeping also the subleading terms on the right hand side of Eq.~\eqref{ap2}, the angular integration for diagram 2 is not infinite but still large, of order $1/\kappa$. Therefore,  we conclude that diagram 4 is suppressed by a factor $ \kappa$ with respect to diagram 2 for $|\vp|\gtrsim \sqrt{a}$ and by a factor $|\vk|^2/a$ for $|\vp|\lesssim \sqrt{a}$.

{\bf Diagram 5}

For the amplitudes represented globally by diagram 5 in Fig.~\ref{fig.dia} there are no enhanced vertices  like those discussed above for diagram 2 or 4. For some kaon arrangements, it is possible  that the intermediate kaon is nearly on-shell but each of the vertices in these amplitudes require one power of small three-momentum. Then, the ratio between the enhanced propagator and the suppressed vertices is ${\cal O}(1)$. As a result, these diagrams are at most of ${\cal O}(g^2/f^2)$.

{\bf Diagram 6}

The enhanced configurations have an outgoing kaon on the leftmost vertex 
and an incoming one on the vertex at the far right. In between the kaon propagators are each of them of  size $1/a$. On the other hand, the coupling of the external vector resonances with the kaons is suppressed by small powers of three-momentum. Then, the size of the amplitudes is estimated to be 
\begin{align}
\frac{g^2}{f^2}\frac{m_K^2}{a}\frac{|\vk|^2}{a}\,.
\end{align}
The suppression compared to the diagram 2 happens in the same way as for diagram 4.
For $|\vp|\lesssim \sqrt{a}$ the last factor in the previous equation is small, and for $|\vp|\gtrsim \sqrt{a}$ there is a suppression due to the angular projection. Indeed, from the vertices involving the vector states one gets the product
\begin{align}
\epsilon(p)\cdot k'_{1,2} \, \epsilon(p')\cdot k_{1,2}\,,
\end{align}
which implies that the following angular structures [see Eq.~\eqref{epsvec}]
\begin{align}
&(\hat{\mathbf{n}}\cdot \vp')^2~,\nn\\
&(\hat{\mathbf{n}}\cdot \vp')(\hat{\mathbf{n}}\cdot \vq)~,\nn\\
&(\hat{\mathbf{n}}\cdot \vq')(\hat{\mathbf{n}}\cdot \vq)~.
\label{str.6}
\end{align}
are present. These terms are multiplied by $1/a^2$ which has a lessened angular dependence since $a$ is $2m_K \delta$ close to threshold. The integrals over $\hat{\vq}$ and $\hat{\vq}'$  in the second and third lines
of Eq.~(\ref{str.6}) are zero. The first line instead is finite and gives a contribution that compared with diagram 2 is suppressed by a factor $\kappa \vp^2 /a$.

{\bf Diagram 7}

A $K\bar{K}$ pair must couple to the vector propagator. In addition, the kaon and anti-kaon in the pair cannot belong both to the same $|K\bar{K}\ra_0$ state because the latter is in S-wave. As a result, the four-momentum running through the vector propagator nearly vanishes due to the vicinity to the $K\bar{K}$ threshold.  Notice that from ${\cal L}_{V\Phi^2}$ in Eq.~(\ref{vvff}) the vertex for a vector resonance coupled to a 
$K\bar{K}$ is proportional to the difference of the four-momenta of the kaon and the anti-kaon. 

\begin{figure}[ht]
\psfrag{Kp}{$K^+$}
\psfrag{Km}{$K^-$}
\psfrag{K0}{$K^0$}
\psfrag{K0b}{$\bar{K}^0$}
\centerline{\epsfig{file=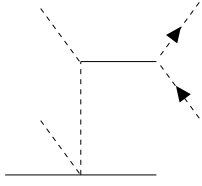,width=.15\textwidth,angle=0}}
\vspace{0.2cm}
\caption[pilf]{\protect \small
Configurations left for the coupling of a $K\bar{K}$  to an intermediate vector resonance. 
\label{fig.7}}
\end{figure}

The diagram shown in Fig.~\ref{fig.7} is obtained by modifying the local lowest order meson-meson chiral vertex in the diagram 2 of Fig.~\ref{fig.dia} by the exchange of a vector resonance between the kaons with vanishing four-momentum transfer.  It is well known that for these modifications to be meaningful~\cite{birse}, they must be calculated together with the exchange of the octet of axial-vector resonances $a_\mu$ in Eq.~(\ref{extvec}). After adding them, the result at the lowest chiral order is not modified i.e. the corrections are of higher order, suppressed by powers of $m_P^2/M_{V,A}^2$ where $m_P$ is  the mass of a pseudoscalar meson and $M_{V,A}$, those of the first octet of vector and axial-vector resonances. Let us stress that in Ref.~\cite{npa} a very good description of the $I=0$,1 meson-meson scattering data was achieved by taking the lowest order Chiral Perturbation Theory amplitudes as the interacting kernels. At the two kaon threshold the $I=0,1$, S-waves  are dominated by the presence of the $f_0(980)$ and $a_0(980)$ resonances, in that order. Both resonances are well reproduced in Ref.~\cite{npa}, with the same approach followed here. Therefore, since we take diagram 2 into account, we  can neglect the contribution of diagram 7 when summed with others not drawn in Fig.~\ref{fig.dia} and that also include the exchange of axial-vector resonances, expressing the result in terms of the full S-wave $K\bar{K}$ strong amplitude derived in Ref.\cite{npa}.

{\bf Diagram 8}

The enhanced configurations in diagram 8 have an outgoing kaon coupled to the vertex at the far left and an incoming one on the next vertex to the right. In this way the intermediate kaon propagator $\sim 1/a$ while the vector resonance propagator is $\sim 1/a'$. The two vertices with two pseudoscalars and one vector resonance involve small external three-momenta. Altogether, this diagram is of order
\begin{align}
\frac{g^2}{f^2}\frac{g^2 f^2 |\vk|^2}{a a'}~.
\label{diag.8}
\end{align}
Numerically, $g^2f^2\simeq 0.4^2$~GeV$^2=0.64\, m_K^2$ which implies already some suppression. In addition there is an additional reduction due to the angular integration, similarly to the situation explained above for diagram 4, so that the additional suppression factor $\kappa$  with respect to diagram 2 applies also here. 

{\bf Diagram 9}

Here, the enhanced configurations have, from left to right, a kaon going out, the next one coming in, another leaving and the last one entering the vertex. This means that the kaon propagators are of size $1/a$ and the intermediate vector resonance propagator is of size $1/a'$. In addition, one has now four vertices involving one vector resonance and two pseudoscalars. Each of them proportional to the difference between two slow kaon four-momenta. Then, the corresponding amplitudes go as
\begin{align}
\frac{g^2}{f^2}\frac{|\vk|^4 g^2f^2}{a^2 a'}~.
\end{align} 
The ratio $|\vk|^4/a^2$ is ${\cal O}(1)$ but for $|\vp|\lesssim \sqrt{a}$ is suppressed. Calculating explicitly the intermediate vertices, attached to the vector meson propagator, a factor $\delta^2$ appears, so that $\delta^2/a$ is suppressed by a factor $\delta/m_K$. Besides, the angular projection suppression due to the vertices involving the external resonances also operates here for their ratio with $a'$. As a result there is an extra factor $\kappa$ compared with diagram 2.

{\bf Diagram 10}

The enhanced configurations arise when there are, from left to right, a kaon leaving the diagram and another entering it on the vertices that couple two pseudoscalars with an external vector resonance. In this case the intermediate kaon propagators are $\sim 1/a$. These contributions are of order
\begin{align}
\frac{g^2}{f^2}\frac{m_K^2}{a}\frac{g^2f^2}{M_\phi^2}\frac{|\vk|^2}{a}~.
\end{align}
The last factor is due to the vertices with the external resonances and involve small external kaon three-momenta. The angular projection suppression operates here similarly as for diagram 6.

In analogy to diagram 7, the diagrams 9 and 10 are vector resonance contributions to the meson-meson scattering vertex. They must be accompanied by other diagrams involving also the exchange of axial-vector resonances so that the final modification of the lowest order chiral amplitude is further suppressed as indicated in the discussion for the diagram 7.

{\bf Diagram 11}

This diagram is similar to diagram 5 but including an extra vector resonance exchange that modifies the four-pseudoscalar one-vector vertex to the right of the diagram 5. As for the latter there is a suppression of the enhanced intermediate propagator, when the external kaon to the left of the diagram is leaving, because it is quadratic in the external small three-momentum. Therefore, it is just ${\cal O}(g^2/f^2)$.

{\bf Diagram 12--17} 

As in Fig.~\ref{fig.7}, one must have a leaving and entering $K\bar{K}$ pair attached to every intermediate vector meson line. As a result, from Eq.~(\ref{nonabe}) one can conclude that all these diagrams are zero. Diagram 12 vanishes because there are no four-vector meson vertices coupling $\phi \phi \rho^0\rho^0$, $\phi \phi \rho^+\rho^-$, $\phi \phi \omega\omega$ and in general  $\phi\phi V W$, with $V$ and $W$ any vector state. Diagram 13 is also zero because there are no three-vector resonance vertices coupling $ \phi\phi V$, with $V$ any vector resonance.  For the same reason diagrams 14 and 15 are also zero. Finally, diagrams 16 and 17 vanish because there are no vertices with three vector resonances that couple $\phi \omega\omega$, $\phi \rho^0\rho^0$ and $\phi\rho^+\rho^-$.

  \end{document}